\documentclass[aps,twocolumn,floatfix]{revtex4-1} 

\usepackage{amsmath,amsfonts,amssymb,graphicx,times,bbm}
\usepackage{dsfont}

\usepackage[usenames,dvipsnames]{xcolor} 
\usepackage{ulem} 

\usepackage{hyperref}

\hypersetup{colorlinks=true,
						linkcolor=red,          
						citecolor=blue,        
						filecolor=magenta,      
						urlcolor=cyan           
}

\begin{document}

\bibliographystyle{apsrev}
\newtheorem{theorem}{Theorem}
\newtheorem{corollary}{Corollary}
\newtheorem{definition}{Definition}
\newtheorem{proposition}{Proposition}
\newtheorem{lemma}{Lemma}
\newcommand{\proofend}{\hfill\fbox\\\medskip }
\newcommand{\proof}[1]{{\bf Proof.} #1 $\proofend$}
\newcommand{\nn}{{\mathbbm{N}}}
\newcommand{\rr}{{\mathbbm{R}}}
\newcommand{\cc}{{\mathbbm{C}}}
\newcommand{\zz}{{\mathbbm{Z}}}
\newcommand{\mbp}{\ensuremath{\spadesuit}}
\newcommand{\je}{\ensuremath{\heartsuit}}
\newcommand{\jd}{\ensuremath{\clubsuit}}
\newcommand{\id}{{\mathbbm{1}}}
\renewcommand{\vec}[1]{\boldsymbol{#1}}
\newcommand{\me}{\mathrm{e}}
\newcommand{\mi}{\mathrm{i}}
\newcommand{\md}{\mathrm{d}}
\newcommand{\sg}{\text{sgn}}

\def\>{\rangle}
\def\<{\langle}
\def\({\left(}
\def\){\right)}

\newcommand{\ket}[1]{|#1\>}
\newcommand{\bra}[1]{\<#1|}
\newcommand{\braket}[2]{\<#1|#2\>}
\newcommand{\ketbra}[2]{|#1\>\!\<#2|}
\newcommand{\proj}[1]{|#1\>\!\<#1|}
\newcommand{\avg}[1]{\< #1 \>}

\renewcommand{\tensor}{\otimes}

\newcommand{\einfuegen}[1]{\textcolor{PineGreen}{#1}}
\newcommand{\streichen}[1]{\textcolor{red}{\sout{#1}}}
\newcommand{\todo}[1]{\textcolor{blue}{(ToDo: #1)}}
\newcommand{\transpose}[1]{{#1}^t}

\newcommand{\magenta}[1]{\textcolor{Magenta}{#1}}


\title{Quantifying entanglement with scattering experiments}

\author{O.\ Marty,$^{1,2}$ M.\ Epping,$^3$  H.\ Kampermann,$^3$ D.\ Bru\ss,$^3$, M.B.\ Plenio,$^{1,2}$ and M.\ Cramer$^{1,2}$}
\affiliation{
$^1$Institut f\"ur Theoretische Physik, Albert-Einstein Allee 11, Universit\"at Ulm, D-89069 Ulm, Germany\\
$^2$Institute for Integrated Quantum Science and Technology, Albert-Einstein Allee 11, Universit\"at Ulm, D-89069 Ulm, Germany\\
$^3$Institute for Theoretical Physics III, Heinrich-Heine-Universit\"at D\"usseldorf, Universit\"atsstr. 1, 40225 D\"usseldorf, Germany
}

\begin{abstract}
We show how the entanglement contained in states of spins arranged on a lattice may be quantified with observables arising in scattering experiments.
We focus on the partial differential cross-section obtained in neutron scattering from magnetic materials but our results are sufficiently general such that they may also be applied to, e.g., optical Bragg scattering from ultracold atoms in optical lattices or 
from ion chains. We discuss resonating valence bond states and ground and thermal states of experimentally relevant models---such as Heisenberg, Majumdar-Ghosh, and XY
models---in different geometries and with different spin numbers. As a by-product, we find that for the one-dimensional XY model in a transverse
field such measurements reveal factorization and the quantum phase transition at zero temperature.
\end{abstract}

\maketitle

\date{\today}


\section{Introduction}
\label{section:introduction}

Entanglement is a key resource for performing quantum information tasks \cite{PlenioV2007,HorodeckiHHH2009}. At low
temperatures, it occurs naturally
in quantum many-body systems and its amount (more concretely, its scaling with the size of system partitions) relates to the complexity
of descriptions of such systems \cite{AudenaertEPW2002,OsborneN2002,Vidal2003,PlenioEDC05,EisertCP2010}. 
It also serves to characterize exotic states of matter, a prominent example being topological spin liquids, see, e.g., the recent Refs.~\cite{JiangWB2012,Depenbrock2012}. 
While the task of merely verifying that
entanglement is present \cite{Guhne2002,Guhne2009} is quite established and has been demonstrated in a number of experiments
\cite{Barbieri2003,Bourennane2004,Wineland6qubits,Haffner8qubits,EsteveGWGONature2008,LerouxSVPRL2010,RiedelBLHSTNature2010,Monz14qubits,
Yao2012,Chauvet2010,Chiuri2010} ,
quantifying its amount rigorously and without any assumptions is a delicate task and has only very recently been experimentally achieved for
a large many-body system of bosons in optical lattices in Ref.~\cite{Cramer2013} (see also, e.g., Ref.~\cite{Wunderlich2011}
 where the entanglement of a small photonic system was quantified using few measurements). Generally speaking, the difficulty increases with the
number of particles carrying the quantum information, i.e., it is especially delicate for large systems for which the
available measurements are usually very limited and very far from being informationally complete (in which case full state
tomography
\cite{VogelR1989,Haffner8qubits,Cramer2010} would be possible). Here, we are interested in such large systems, namely a large
number of spins arranged on a lattice. In order to quantify the amount of entanglement that is shared between
the spins, we rely only on global measurements typically obtained in scattering experiments.
We achieve this by generalizing results of the recent Refs.\ \cite{KrammerKBBKM2009,CramerPW2011} to
arbitrary spin and to more general observables. In the case of neutron scattering from magnetic materials, this
enables us to quantify entanglement for arbitrary lattice geometries relying solely on the Fourier transform of the scattering cross-section (or, alternatively, measurements that do not resolve the energy of the scattered neutrons). 
Our strategy adopts a principle from quantum information theory that is simple yet powerful
\cite{HorodeckiHH99,Brandao2005,AudenaertP2006,Guehne2007,EisertBA2007}:  Given certain observables and their experimentally obtained expectation values, we ask
what is the minimal amount of entanglement that is consistent with the obtained outcomes, i.e., given
the expectation values of the observables, we minimize over all density matrices that are consistent with them. In this
way, we arrive at the least
amount of entanglement that is consistent with the measurement outcomes and thus we put a lower bound on the
entanglement contained in the sample on
which the measurements were performed. By the very nature of this principle, we need not make {\it any} assumptions on the system (such as, e.g., the
temperature, details of external potentials, the Hamiltonian governing the system, or even the system being in equilibrium).

We consider observables that arise in scattering experiments from $N$ spins arranged on a lattice. Examples include optical
Bragg scattering from ultracold atoms in optical lattices \cite{Corcovilos2010} or 
from ion chains \cite{Macchiavello2013} and neutron scattering from magnetic materials \cite{Zaliznyak}. These observables may be written as
\begin{equation}
\label{eq:Sq}
\hat{S}(\vec{q})=\sum_{\alpha,\beta}M_{\alpha,\beta}(\vec{q})\hat{S}_{\alpha,\beta}(\vec{q}),
\end{equation}
where, usually, $\vec{q}=\vec{k}_f-\vec{k}_i$ is the scattering vector, i.e. the difference between the final and the initial wave vector.
Here, 
\begin{equation}
\label{equation:generalSq}
\hat{S}_{\alpha,\beta}(\vec{q})=\sum_{i,j=1}^Nf_{i,\alpha}^*(\vec{q})f_{j,\beta}(\vec{q})
\me^{\mi\vec{q}(\vec{r}_i-\vec{r}_j)}
\hat{S}_i^\alpha\hat{S}_j^\beta,
\end{equation}
where $\vec{r}_i$ is the position of the $i$'th spin with corresponding spin operators $\hat{S}_i^\alpha$, $\alpha=x,y,z$, and spin quantum number $s$, and the coefficients  $M_{\alpha,\beta}(\vec{q})$ and $f_{i,\alpha}(\vec{q})$ depend on the system under consideration. While keeping our results as general as possible, we will focus on neutron scattering experiments, in which such observables arise as follows.

The neutrons interact magnetically with the atoms of the target sample, whose magnetic moments mostly originate from the orbital motion and spins of unpaired electrons. In many cases an effective spin
value can be assigned to either the magnetic atoms or to the entire unit-cell \cite{Zaliznyak}. With the formalism
introduced by Van Hove in Ref.~\cite{VanHove1954}, the partial differential cross-section can be expressed in terms of
time-dependent correlation functions. Accordingly, for unpolarized neutrons, the magnetic cross-section is proportional to \cite{Zaliznyak}
\begin{equation}
\begin{split}
\label{eq:cross-section} 
\frac{k_f}{k_i}\sum_{\alpha,\beta}(\delta_{\alpha,\beta}-\bar{q}_\alpha\bar{q}_\beta)\sum_{i,j}f_{i,\alpha}(\vec{q})^*f_{j,\beta}(\vec{q}
)\me^{\mi\vec{q}(\vec{r}_i-\vec{r}_j)}
\hspace{0.5cm}\\
\times\int\md t\,\me^{-\mi\omega t}\langle\hat{S}_i^\alpha\hat{S}_j^\beta(t)\rangle,
\end{split}
\end{equation}
where $\omega$ is the energy transferred to the sample. Furthermore, 
$f_{i,\alpha}(\vec{q})=F_i(\vec{q})g_{i,\alpha}$, where $F_i(\vec{q})$ and $g_{i,\alpha}$ denote the magnetic form factor and
the Land\'e factor of the $i$'th site, respectively, and $\bar{\vec{q}}=\vec{q}/|\vec{q}|$. In general, we allow the g-factor
to be anisotropic and
$f_{i,\alpha}(\vec{q})$ to be site-dependent, where $i$ labels the lattice sites with corresponding effective values of $f_{i,\alpha}$  and
$\hat{S}_i^\alpha$ (corresponding to an effective spin quantum number $s$). The magnetic form factor $F_i(\vec{q})$ stems from the finite
extent of the electron orbitals seen by the neutron with wavelength of the order of interatomic distances. To determine it, a detailed
knowledge about the electronic wave functions of the magnetic atoms in the scatterer is required, and its values
may be found in the
literature.
As $k_{i,f}$ are known, one may multiply (\ref{eq:cross-section}) by $k_i/k_f$ and take the Fourier transform to obtain the instantaneous
scattering function $S(\vec{q})=\langle\hat{S}(\vec{q})\rangle$, where $\hat{S}(\vec{q})$ is as in Eq.~(\ref{eq:Sq}) with $M_{\alpha,\beta}(\vec{q})= \delta_{\alpha,\beta}-\bar{q}_\alpha\bar{q}_\beta$.
Alternatively, $S(\vec{q})$ may be obtained if the requirements of the static approximation are fulfilled \cite{static_approx}
and the final energy is not resolved. For quasi-one- or two-dimensional systems one may also consider a special scattering geometry
\cite{BirgeneauSS1971,BirgeneauGKLWEYS1999,KimBCGKLWAEKHES2001} to obtain $S(\vec{q})$.

In Section~\ref{section:results}, we show how a lower bound to the entanglement shared among $N$ spins---as
quantified in terms of the {\it best separable approximation} \cite{LewensteinS1998} or the {\it (generalized) robustness of entanglement}
\cite{VidalT1999,Steiner2003}---may be obtained from the expectation value of $\hat{S}(\vec{q})$ in Eq.~(\ref{eq:Sq}). In this way, we
quantify entanglement of a collection of $N$ spins without any assumption on the system. In Sections~\ref{section:numerics} and
\ref{section:analytics} we show that our method allows to quantify the entanglement of ground and thermal states corresponding to several model Hamiltonians. We
conclude with a summary and outlook in Section~\ref{section:summary}.

\section{Main results}
\label{section:results}

In this section we will show how observables as in Eq.\ \eqref{eq:Sq} may serve as lower bounds to the entanglement. We will consider several entanglement monotones and a particular simple form will be derived for the best separable approximation $\mathcal{E}_{BSA}[\hat{\varrho}]$ in the neutron scattering setting: 
For any scattering vector $\vec{q}$, we find (see below and Appendix~\ref{appendix:bounds} for details)
\begin{equation}
\mathcal{E}_{BSA}[\hat{\varrho}]\ge
1-\frac{1}{c_{\min}}\sum_{\alpha,\beta}(\delta_{\alpha,\beta}-\bar{q}_\alpha\bar{q}_\beta)\langle\hat{S}_{\alpha,\beta}(\vec{q})\rangle,
\end{equation}
where $c_{\min}$ is a constant that depends on the spin quantum number $s$ and the magnetic form factors $F_i(\vec{q})$ and Land\'e factors $g_{i,\alpha}$.
Hence, a measurement of the Fourier transform of the magnetic scattering cross-section at a single scattering vector directly provides a lower bound to the entanglement contained in the sample. A numerical analysis of the above bound may be found in Section~\ref{section:numerics} (see Figs.~\ref{fig:1d}-\ref{fig:2d}) for different physical models that describe, among others, the magnetic compounds summarized in Table~1.

In the remainder of this section, we detail the derivation of the above bound and the bounds on robustness of entanglement measures. We start with a detailed description of the scattering observables under consideration.

\subsection{The observables under consideration}
We will see below that for many systems, a measurement of $\langle\hat{S}(\vec{q})\rangle$ at a single scattering vector $\vec{q}$ suffices to put meaningful tight lower bounds on
the entanglement quantified via the best separable approximation (BSA). For the
robustness measures, however, we have found that measurements at a single scattering vector
$\vec{q}$ do not suffice to obtain non-trivial bounds for large systems (see also Ref.\ \cite{KrammerKBBKM2009}). To this end,
we incorporate knowledge of $\langle\hat{S}(\vec{q})\rangle$ at several $\vec{q}$ by slightly generalizing the observables in the
introduction to observables of the form
\begin{equation}
\label{eq:summation}
\hat{S}=\sum_{\vec{q}\in Q}\hat{S}(\vec{q}).
\end{equation}
As we will see, this summation over measurements obtained at several scattering vectors will result in positive entanglement
bounds even in the thermodynamic limit. Here, $Q\subset\rr^3$ is some collection of scattering vectors and
$\hat{S}(\vec{q})$ is defined as in Eq.~(\ref{eq:Sq}), where we make the following
assumptions on the coefficients $M_{\alpha,\beta}(\vec{q})\in\cc$ and $f_{i,\alpha}(\vec{q})\in\cc$: We assume that
the $3\times 3$ matrix $M(\vec{q})$ with entries $M_{\alpha,\beta}(\vec{q})$ is Hermitian, i.e., 
$M_{\alpha,\beta}(\vec{q})=M^*_{\beta,\alpha}(\vec{q})$, and positive semi-definite. We further assume that for each $i=1,\dots,N$ the
$3\times 3$ matrix $M^{(i)}$ with entries
\begin{equation}
\label{equation:Mi}
M^{(i)}_{\alpha,\beta}=\sum_{\vec{q}\in Q}f_{i,\alpha}^*(\vec{q})f_{i,\beta}(\vec{q})M_{\alpha,\beta}(\vec{q})
\end{equation}
is real and symmetric, i.e., $M^{(i)}_{\alpha,\beta}=M^{(i)}_{\beta,\alpha}\in\rr$. All these assumptions are
fulfilled, e.g., in the neutron scattering setting, for which we have 
$M(\vec{q})=\id-\bar{\vec{q}}\bar{\vec{q}}^t$ (see Eq.~(\ref{eq:cross-section}))  and
$f_{i,\alpha}(\vec{q})=F_i(\vec{q})g_{i,\alpha}$ with $g_{i,\alpha}\in\rr$.

\subsection{Lower bounds to the entanglement}
In what follows, we consider multipartite entanglement in the following sense. Every state $\hat{\varrho}$ that is not fully separable, i.e., of the form
\begin{equation}
\label{sep}
\sum_np_n\bigotimes_{i=1}^{N} \hat{\varrho}^{(n)}_i\in \mathcal{S},
\end{equation}
with $p_n>0$ and $\sum_n p_n =1$,  will be called entangled. Here, we denoted the set of separable states by $\mathcal{S}$. The
degree of entanglement is then quantified using entanglement monotones \cite{PlenioV2007,HorodeckiHHH2009},
that is, functionals $\mathcal{E}[\hat{\varrho}]$ that do not increase under local operations and classical
communication. 
The monotones under consideration are part of a larger family of monotones that may be expressed as
\cite{Brandao2005}
\begin{equation}
\label{fb}
\mathcal{E}_{\mathcal{C}}[\hat{\varrho}]=-\min_{\hat{W}\in\mathcal{W}\cap\mathcal{C}}\text{tr}[\hat{W}\hat{\varrho}]
\end{equation}
with the convention that $\mathcal{E}_{\mathcal{C}}[\hat{\varrho}]=0$ if the minimization results in a positive number.
Here, $\mathcal{W}$ is the set of {\it entanglement witnesses} (Hermitian operators with non-negative expectation
value for every separable state, i.e.,$\langle\hat{W}\rangle_{\text{sep.}}\ge 0$, see
Ref.~\cite{Guhne2009} for a review) and the set $\mathcal{C}$ depends on the chosen
entanglement measure: If
\begin{equation}
\label{C_BSA}
\mathcal{C}=\bigl\{\hat{W}\in\mathcal{W}\,\big|\,\id+\hat{W}\ge 0\bigr\}
\end{equation}
then $\mathcal{E}_{\mathcal{C}}[\hat{\varrho}]=\mathcal{E}_{BSA}[\hat{\varrho}]$ quantifies entanglement in terms of the best separable approximation \cite{LewensteinS1998}, which, 
in essence, answers the question of how much of a separable state is contained in the state $\hat{\varrho}$. For
\begin{equation}
\label{C_R}
\mathcal{C}=\bigl\{\hat{W}\in\mathcal{W}\,\big|\,\text{tr}[\hat{W}\hat{\sigma}]\le 1 \,\forall\, \hat{\sigma}\in\mathcal{S}\bigr\}
\end{equation}
we have $\mathcal{E}_{\mathcal{C}}[\hat{\varrho}]=\mathcal{E}_{R}[\hat{\varrho}]$, quantifying entanglement in terms of the robustness of entanglement. Finally, if
\begin{equation}
\label{C_GR}
\mathcal{C}=\bigl\{\hat{W}\in\mathcal{W}\,\big|\, \id-\hat{W}\ge 0\bigr\}
\end{equation}
then $\mathcal{E}_{\mathcal{C}}[\hat{\varrho}]=\mathcal{E}_{GR}[\hat{\varrho}]$ is the generalized robustness of entanglement. These robustness measures
\cite{VidalT1999,Steiner2003} quantify the minimal amount of noise (in the form of general state in the case of the generalized robustness and in the form of a separable state in the case of the robustness) that must be mixed in to make $\hat{\varrho}$ separable.

Instead of minimizing over all the entanglement witnesses $\hat{W}\in\mathcal{W}\cap\mathcal{C}$, we construct a single member of the set
$\mathcal{W}\cap\mathcal{C}$ of the form
\begin{equation}
\label{ab}
 \hat{W}_{\hat{S},\mathcal{C}} = a_{\mathcal{C}} \hat{S} + b_{\mathcal{C}} \id
\end{equation}
with appropriate real coefficients $a_{\mathcal{C}}$ and $b_{\mathcal{C}}$ (which will depend on the set of scattering vectors $Q$ and the matrices $M(\vec{q})$) and $\hat{S}$ as in the previous section. By inspection of Eq.~(\ref{fb}), we see that any $\hat{W}\in\mathcal{W}\cap\mathcal{C}$ gives a lower bound to the entanglement monotone
and thus for any state $\hat{\varrho}$, one has
\begin{equation}
\mathcal{E}_{\mathcal{C}}[\hat{\varrho}]\geq -a_{\mathcal{C}}\langle\hat{S}\rangle-b_{\mathcal{C}}, \label{eq:lowerboundtomonotone}
\end{equation}
which depends only on the expectation value $\langle\hat{S}\rangle=\text{tr}[\hat{S}\hat{\varrho}]$. The coefficients are found in the
following way. As the matrices $M(\vec{q})$ are assumed to be positive semidefinite, it is straightforward to show that $\hat{S}$ is also
positive semidefinite, see Appendix~\ref{appendix:bounds}. Furthermore, one may derive bounds on the minimal and maximal
achievable expectation value in fully separable states
\begin{equation}
\label{eq:boundsep}
c_{\min}\le \langle\hat{S}\rangle_{\text{sep.}} \le c_{\max}.
\end{equation}
Together with positive semidefiniteness of $\hat{S}$, such bounds allow us to arrive at witnesses that are of the form as in Eq.~(\ref{ab})
and members of the set $\mathcal{W}\cap\mathcal{C}$. One readily verifies that the coefficients
\begin{equation}\label{eq:coefficientsab}
\begin{split}
a_{BSA}&=\frac{1}{c_{\min}},\;\;\;b_{BSA}=-1,\\
a_{R}&=-\frac{1}{c_{\max}-c_{\min}},\;\;\;b_{R}=-c_{\max}a_{R},\\
a_{GR}&=-\frac{1}{c_{\max}},\;\;\;b_{GR}=1,\\
\end{split}
\end{equation}
fulfil the necessary requirements as defined in Eqs.~(\ref{C_BSA}-\ref{C_GR}). It remains to make the bounds $c_{\min}$ and
$c_{\max}$ explicit. 

\begin{figure*}[!t]
	\begin{center}
		\includegraphics[width=0.99\textwidth]{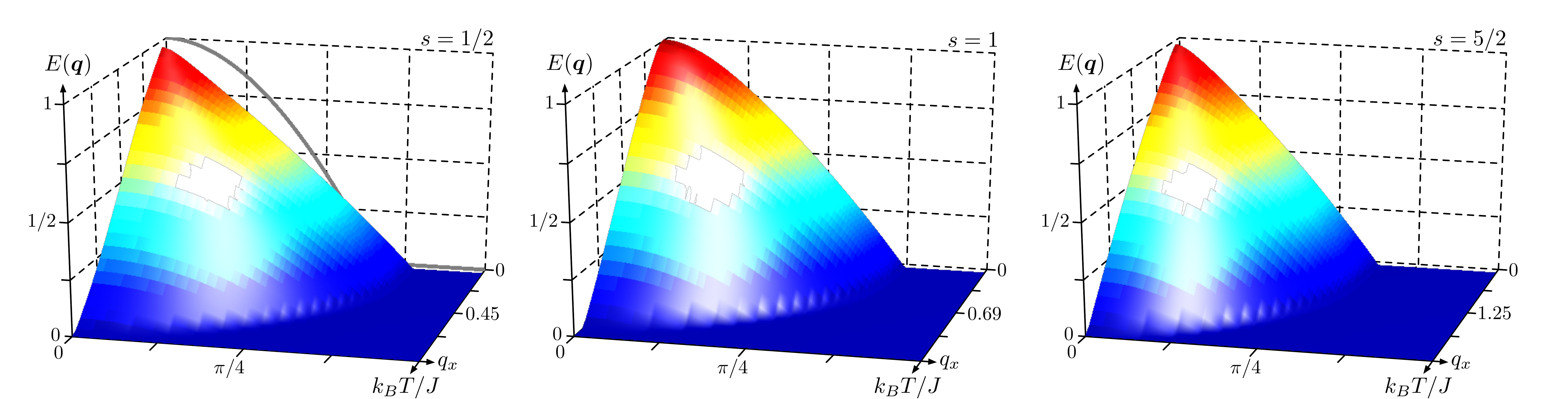}
	\end{center}
	\caption{Lower bound $E_{BSA}[\hat{\varrho}](\vec{q})=E_{BSA}[\hat{\varrho}](q_x)$ (Eq.~(\ref{equation:BSAbound})) to the entanglement $\mathcal{E}_{BSA}[\hat{\varrho}]$ vs. temperature for thermal states of the
quasi-one-dimensional Heisenberg model Eqs.~(\ref{hamiltonian:effD},\ref{hamiltonian:heisenberg1D}) 
	for $s=1/2, 1, 5/2$ (left to right) and $L=900$ \cite{finite_size_footnote}. 
	The gray solid line in the left-most plot depicts the entanglement bound for a ground state of the Majumdar-Gosh model
in the limit
$L\rightarrow\infty$. Note that for all shown models the bound $\mathcal{E}_{BSA}[\hat{\varrho}]\le 1$, which holds for any state $\hat{\varrho}$, is
attained at $T=0$, $q_x=0$.}
	\label{fig:1d}
\end{figure*}

\subsubsection{Lower bound to the best separable approximation}
\begin{figure*}[!t]
	\begin{center}
		\includegraphics[width=2\columnwidth]{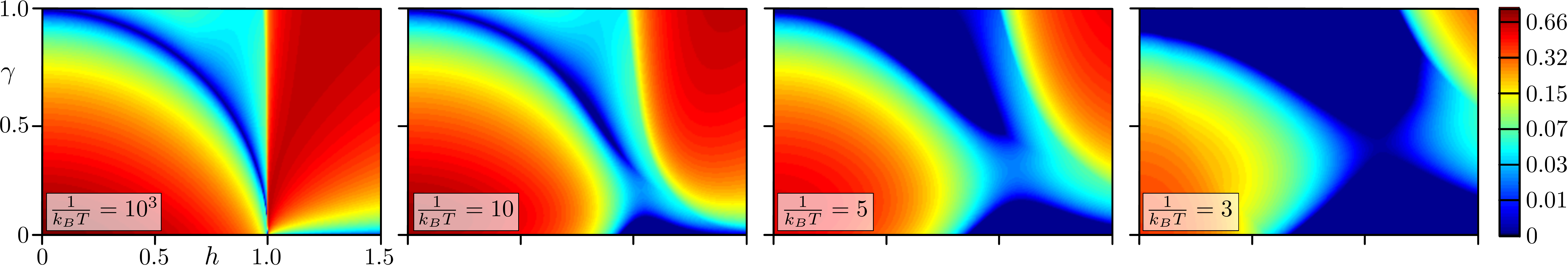}
	\end{center}
	\caption{Lower bound $E_{BSA}[\hat{\varrho}]$ (Eq.~(\ref{equation:BSAbound})) to the best separable approximation
$\mathcal{E}_{BSA}[\hat{\varrho}]$  for thermal
states of a system of
mutually uncoupled chains, Eq.~(\ref{hamiltonian:effD}), each of which is described by the XY-model in Eq.~(\ref{hamiltonian:XY}). Linear dimension is $L=200$ and
the depicted bounds are obtained by optimizing $E_{BSA}[\hat{\varrho}](\vec{q})$ over certain $\vec{q}$ and over the orientation of the chains (see main text). For low
temperature, the phase boundary and factorization circle $\gamma^2+h^2=1$ are clearly visible. Note also that for higher temperature, there are regions on
this circle with finite entanglement. 
	\label{fig:xyBSA}}
\end{figure*}

For each $i=1,\dots, N$, denote the eigenvalues of the $3\times 3$ matrix $M^{(i)}$ in Eq.~(\ref{equation:Mi}) by $m_\alpha^{(i)}$. 
For product states, so
for each summand in Eq.~\eqref{sep}, the expectation value $\langle \hat{S}_i^\alpha \hat{S}_j^\beta \rangle$ can be written as the product $\langle
\hat{S}_i^\alpha\rangle \langle \hat{S}_j^\beta \rangle$ for lattice sites $i\neq j$. The resulting expression can then be
bounded with the help of the eigenvalues of the coefficient matrices $M^{(i)}$ in the following way (see Appendix~\ref{appendix:bounds} for details).
\begin{equation}
\label{eq:mini}
c_{\min}=\sum_{i=1}^N\min_{|\psi\rangle}\sum_\alpha m^{(i)}_\alpha\bigl(\langle\psi|\hat{S}^2_\alpha|\psi\rangle-\langle\psi|\hat{S}_\alpha|\psi\rangle^2\bigr),
\end{equation}
where $\hat{S}_\alpha$, $\alpha=x,y,z$, are the spin operators for a single spin. 
For each $i$, the minimization over pure states $|\psi\rangle\in\cc^{2s+1}$
 may be solved numerically. For some special cases, it may be given explicitly: E.g., for $f_{i,\alpha}(\vec{q})=f(\vec{q})$ and
 $M_{\alpha,\beta}(\vec{q})=\delta_{\alpha,\beta}/|f(\vec{q})|^2$ (similar observables were considered in
\cite{CramerPW2011}), one finds
\begin{equation}
c_{\min}=N|Q|s,
\end{equation} 
where we recall that $s$ is the spin quantum number corresponding to the $\hat{S}_i^\alpha$ and $|Q|$ denotes the number of scattering vectors in the set $Q$. If $f_{i,\alpha}(\vec{q})=f(\vec{q})$, if $M(\vec{q})=(\id-\bar{\vec{q}}\bar{\vec{q}}^t)/|f(\vec{q})|^2$
as in the neutron scattering setting, and if $Q$
contains only one scattering vector, we have 
\begin{equation}
c_{\min}=NC_s,\label{eq:cminneutrononeqindepf}
\end{equation}
where \cite{HofmannT2003}
\begin{equation}
C_s=\begin{cases}
\frac{1}{4} & \text{ for } s=\frac{1}{2},\\
\frac{7}{16} & \text{ for } s=1,
\end{cases}
\end{equation}
and further values are listed in Ref.~\cite{HePDR2011}. The latter yields the following bound on the best separable approximation. For each
$\vec{q}\in\rr^3$, inserting Eq.~(\ref{eq:cminneutrononeqindepf}) into
Eqs.~(\ref{eq:lowerboundtomonotone}) and (\ref{eq:coefficientsab}) leads to
\begin{equation}
\label{equation:BSAbound}
\begin{split}
\mathcal{E}_{BSA}[\hat{\varrho}]&\ge 1-\sum_{\alpha,\beta}\tfrac{\delta_{\alpha,\beta}-\bar{q}_\alpha\bar{q}_\beta}{NC_s}\sum_{i,j}
\me^{\mi\vec{q}(\vec{r}_i-\vec{r}_j)}
\langle\hat{S}_{i}^\alpha\hat{S}_{j}^\beta\rangle\\
&=:E_{BSA}[\hat{\varrho}](\vec{q}).
\end{split}
\end{equation}
Note that this is a general bound for {\it any} state. Whenever the expectation value $E_{BSA}[\hat{\varrho}](\vec{q})$ is accessible, it
provides
a lower bound to the entanglement contained in $\hat{\varrho}$ -- no matter what the underlying Hamiltonian of the system or the
temperature 
might be, no matter whether the system is in equilibrium or not. If, depending on the
experimental situation, $E_{BSA}[\hat{\varrho}](\vec{q})$ is not accessible, i.e. the special form of $M(\vec{q})$ and
$f_{i,\alpha}(\vec{q})$ is not given, one has to use the observable given in Eqs.~(\ref{eq:Sq}) and (\ref{equation:generalSq})
and the general bound in Eq.~(\ref{eq:mini}) needs to be applied.
Note that, for any state, $\mathcal{E}_{BSA}[\hat{\varrho}]\le 1$, i.e., whenever we find $E_{BSA}[\hat{\varrho}](\vec{q})=1$, the bound is in
fact equal to the exact entanglement. In Section~\ref{section:numerics}, we present $E_{BSA}[\hat{\varrho}](\vec{q})$ for several
numerically simulated states, see Figs.~\ref{fig:1d}-\ref{fig:2d}, and in Section~\ref{section:analytics}, we discuss some examples for which $E_{BSA}[\hat{\varrho}](\vec{q})$
may be obtained analytically.

\subsubsection{Lower bound to robustness measures}

The derivation of the general bound may be found in Appendix~\ref{appendix:bounds}, for clarity, we only state it here for the following special case.
We let $f_{i,\alpha}(\vec{q})=f(\vec{q})$ and $M_{\alpha,\beta}(\vec{q})=\delta_{\alpha,\beta}/|f(\vec{q})|^2$ such that our
observable reads
\begin{equation}
\label{eq:obsRob}
\hat{S}=\sum_{\vec{q}\in Q}\sum_{i,j=1}^N
\me^{\mi\vec{q}(\vec{r}_i-\vec{r}_j)}
\sum_{\alpha}
\hat{S}_i^\alpha\hat{S}_j^\alpha.
\end{equation}
We further assume that the $N=N_1N_2N_3$ spins are arranged on a finite three-dimensional Bravais lattice with primitive vectors
$\vec{a}_d$, $d=1,2,3$, such that
$\vec{r}_i=\sum_{d=1}^3i_d\vec{a}_d$ with $i_d\in\{1,\dots,N_d\}$.
Further we assume that 
\begin{equation}
Q\subset\Bigl\{\sum_{d=1}^3q_d\vec{b}_d\,\big|\, q_d\in\frac{1}{N_d}\{0,\dots,N_d-1\}\Bigr\}=:\mathcal{Q},
\end{equation}
where the $\vec{b}_d$ are the reciprocal primitive  vectors.
The upper bound is derived in appendix~\ref{appendix:bounds} and reads
\begin{equation}
\label{eq:cmax}
c_{\max}=N|Q|s+N^2s^2,
\end{equation}
see Eq.~(\ref{eq:upperboundsimple}). 
Hence, whenever the expectation of the observable in Eq.~(\ref{eq:obsRob}) may be obtained, we have the following lower bounds to the robustness measures for {\it any} state.
For all $Q\subset \mathcal{Q}$, we have
\begin{equation}
\label{equation:Rbound}
\begin{split}
\mathcal{E}_{R}[\hat{\varrho}]&\geq \frac{\langle\hat{S}\rangle-N|Q|s}{N^2s^2} -1=:E_R[\hat{\varrho}],\\
\mathcal{E}_{GR}[\hat{\varrho}]&\ge \frac{\langle\hat{S}\rangle}{N|Q|s+N^2s^2}-1=:E_{GR}[\hat{\varrho}].
\end{split}
\end{equation}
We present $E_R[\hat{\varrho}]$ and $E_{GR}[\hat{\varrho}]$ for several
numerically simulated states in Section~\ref{section:numerics}, see Fig.~\ref{fig:rob}, and discuss some analytic examples in Section~\ref{section:analytics}.

\section{Numerical analysis of magnetic materials}
\label{section:numerics}
For all our numerical examples we assume that  $f_{i,\alpha}(\vec{q})=f(\vec{q})$, that the $N=L^3$ spins are arranged on a simple cubic lattice with $\vec{r}_i=\vec{i}\in\{1,\dots,L\}^{\times 3}$ and periodic boundary conditions,
and that
\begin{equation}
Q\subset 2\pi\{0,\dots,L-1\}^{\times 3}/L.
\end{equation}

We will consider ground and thermal states, $\hat{\varrho}=\me^{-\hat{H}/(k_BT)}/Z$, of quasi-one- and two-dimensional
Hamiltonians, that is,
Hamiltonians of the form
\begin{equation}
\label{hamiltonian:effD}
\hat{H}=\sum_{i_z,i_y=1}^N\hat{H}_{1D}^{(i_z,i_y)}\;\;\text{or}\;\;\hat{H}=\sum_{i_z=1}^N\hat{H}_{2D}^{(i_z)},
\end{equation}
i.e., Hamiltonians that correspond to $L^2$ mutually uncoupled chains or Hamiltonians that correspond to $L$
mutually uncoupled two-dimensional systems.
We further assume that the individual chains are governed by the same one-dimensional Hamiltonian $\hat{H}_{1D}$ and will give numerical
examples for the one-dimensional Heisenberg model and the $XY$-chain. Similarly, we assume that
the individual two-dimensional systems are governed by the same $\hat{H}_{2D}$ and provide numerical examples for it being the
two-dimensional Heisenberg model.

Results for thermal states are obtained using the loop algorithm of the ALPS quantum Monte Carlo library \cite{alps}. For
details on the simulation of effective one- and two-dimensional models and the symmetries of the models under consideration see Appendix~\ref{appendix:effectiveD} and \ref{appendix:symmetries}.

We start with quasi-one-dimensional models, the first of which is the antiferromagnetic one-dimensional Heisenberg model, i.e., the individual chains are governed by the Hamiltonian
\begin{equation}
\label{hamiltonian:heisenberg1D}
\hat{H}^H_{1D}=J\sum_{\langle i,j\rangle}\hat{\vec{S}}_{i}\cdot\hat{\vec{S}}_{j}=
J\sum_{\langle i,j\rangle}\sum_\alpha\hat{S}^\alpha_{i}\hat{S}^\alpha_{j},
\end{equation}
where $\langle\cdot,\cdot\rangle$ denotes summation over nearest neighbors. Various materials may approximately be described by
such mutually uncoupled chains and have been studied experimentally using neutron
scattering, see Table~1 for some examples. In Fig.~1, we present results for the lower bound $E_{BSA}[\hat{\varrho}](\vec{q})$, which, due to symmetries of the considered model, is independent of $q_y$ and $q_z$  (see Appendix~\ref{appendix:symmetries} for
details).

\begin{table}
\begin{center}
\begin{tabular}{|c|c|c|c|c|c|}
\hline
Compound &Effective $D$&$s$&$J $ [K]& studied at $T$ [K]& in Ref.\\
\hline\hline
Cs$_2$CuCl$_4$ & $1$&1/2& $ 4$ &$0.06$ &\cite{ColdeaTCMDT1997}\\
\hline
CsNiCl$_3$ &$1$& 1 & $17$ & $1.6$ &\cite{SteinerKKPP1987}\\
\hline
CFTD &2& $1/2$&$72$ &$1.5$ &\cite{ChristensenRMHPECRA2007}\\
\hline
La$_2$CuO$_4$ &$2$&$1/2$ & $1567$ & $337$ & \cite{BirgeneauGKLWEYS1999}\\
\hline
SrCuO$_4$Cl$_2$ &$2$&$1/2$& $1451$ &$10$ &\cite{KimBCGKLWAEKHES2001}\\
\hline
K$_2$NiF$_4$ & $2$&$1$& $112$ & $4.2$ &\cite{BirgeneauSS1971}\\
\hline 
Rb$_2$MnF$_4$ & $2$& $5/2$ &$ 8$ & $10$ &\cite{LeeGWBS1998}\\
\hline
\end{tabular}
\end{center}
\caption{Various materials that have been studied using neutron scattering and that may approximately be described
by quasi-one- or
two-dimensional Heisenberg Hamiltonians.}
\end{table}

\begin{figure*}[t!]
	\begin{center}
		\includegraphics[width=0.99\textwidth]{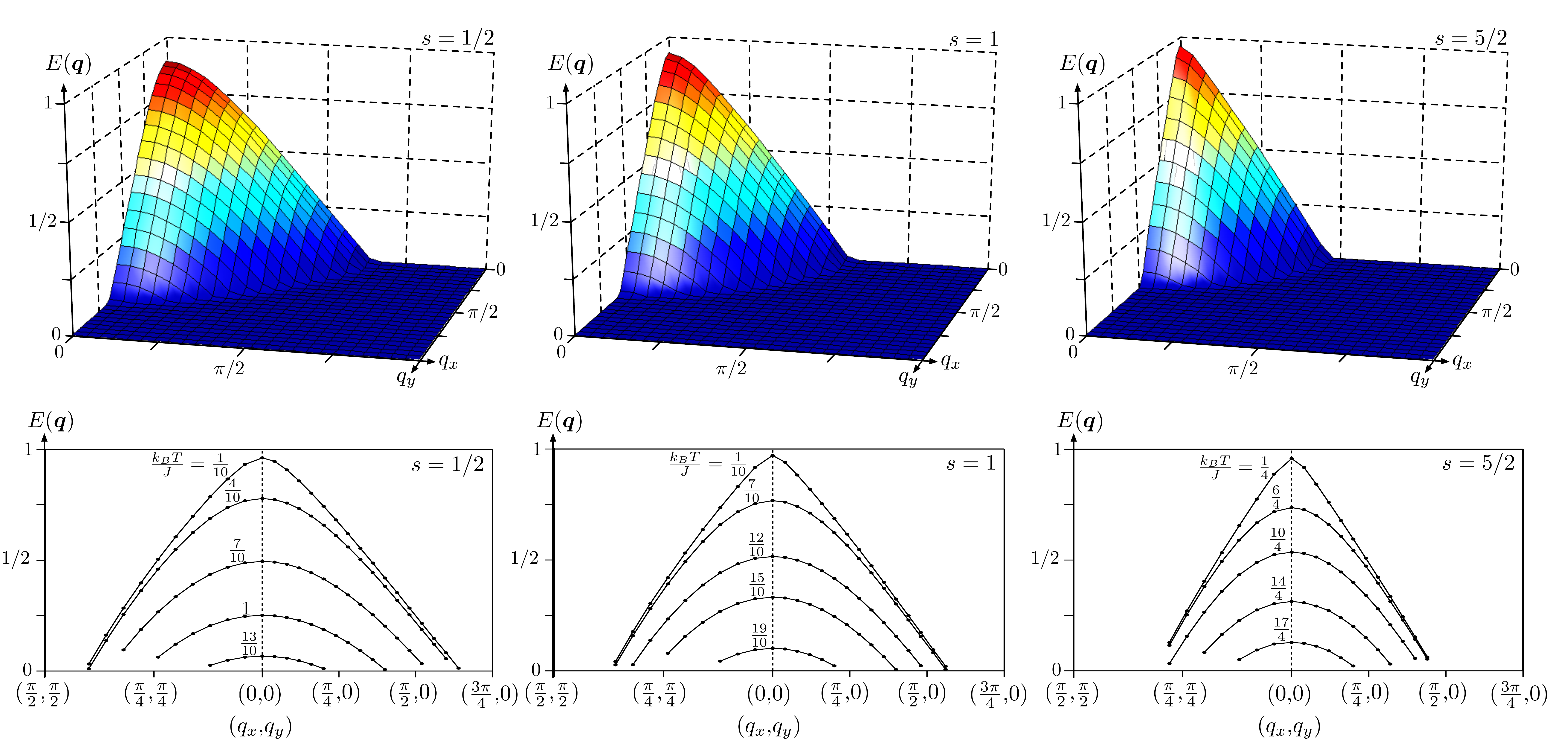}
	\end{center}
	\caption{Lower bound $E_{BSA}[\hat{\varrho}](\vec{q})=E_{BSA}[\hat{\varrho}](q_x,q_y)$ (Eq.~(\ref{equation:BSAbound})) to the entanglement $\mathcal{E}_{BSA}[\hat{\varrho}]$ for thermal states  of the quasi-two-dimensional Heisenberg model
  in Eq.\ (\ref{hamiltonian:heisenberg2D}) 
	with $s=1/2, 1, 5/2$ (left to right). Top row shows $E_{BSA}[\hat{\varrho}](q_x,q_y)$ for $T/J=1/4$, bottom row shows cuts through the first Brillouin zone
for different temperatures. Cuts are along the line from  $(q_x,q_y)=(\pi/2,\pi/2)$ to $(0,0)$ and and along the $x$-axis from $(0,0)$ to
$(3\pi/4,0)$. Simulated system size is $L=30$ \cite{finite_size_footnote}. 
	Lines are guides to the eye and only data points with $E_{BSA}[\hat{\varrho}](q_x,q_y)>0$ and $(q_x,q_y)\in 2\pi\{0,\dots,L-1\}^{\times 2}/L$ are shown.}
	\label{fig:2d}
\end{figure*}

As a second quasi-one-dimensional example, we consider the 
spin-$1/2$ XY-chain in a transverse magnetic field,
\begin{equation}
\label{hamiltonian:XY}
\hat H^{XY} _{1D}= \sum_{\langle i,j\rangle} \bigl[(1+\gamma) \hat S_{i}^x \hat S_{j}^x + (1-\gamma) \hat S_{i}^y \hat S_{j}^y \bigr] - h
\sum_{i} \hat S_i^z,
\end{equation}
where $\gamma$ is the anisotropy parameter and $h$ denotes the magnetic field. The system undergoes a quantum phase transition
at the
critical value $h=1$ and the ground state factorizes for $\gamma^2+h^2 = 1$.
See Ref.~\cite{BasakC1989} for a comparison of this model to experimental data on Cs$_2$CoCl$_4$ and Ref.~\cite{GoovaertsDRS1984}
for confirmation of the one-dimensional spin-$1/2$ XY character of the interactions between the pseudospins of the Pr$^{3+}$ ions in
PrCl$_3$.
The spin-correlation functions for thermal states of this model were extensively studied by Barouch and McCoy in \cite{BarouchMcC1971} and
may be obtained numerically for very large chain lengths. We present lower
bounds to the best separable approximation of thermal states of this model in Fig.~2. These bounds are obtained by maximizing the bound $E_{BSA}[\hat{\varrho}](\vec{q})$ over all
$\vec{q}\in 2\pi\{0,\dots,L-1\}^{\times 3}/L$ with $\vec{q}\ne\vec{0}$ and $\bar{q}_x\bar{q}_y=0$. Further, we maximize
over three possible orientations of the chains in real space (oriented along the $x$, $y$, or $z$ direction),
see Appendix \ref{appendix:symmetries} for details.

Finally, in Fig.~\ref{fig:2d}, we present results for the quasi-two dimensional model, in which each two-dimensional subsystem is governed by
the Heisenberg model such that the total Hamiltonian reads
\begin{equation}
\label{hamiltonian:heisenberg2D}
\hat{H}^H=J\sum_{\langle \vec{i},\vec{j}\rangle}\delta_{i_z,j_z}\sum_\alpha\hat{S}^\alpha_{\vec{i}}\hat{S}^\alpha_{\vec{j}},
\end{equation}
where we recall that $\vec{i}=(i_x,i_y,i_z)\in\{1,\dots,L\}^{\times 3}$.
Due to the symmetries of this model, the bound $E_{BSA}[\hat{\varrho}](\vec{q})$ in Eq.~(\ref{equation:BSAbound}) is independent of $q_z$,
see Appendix~\ref{appendix:symmetries} for details. For compounds well described by the quasi-two-dimensional Heisenberg model see Table~1.

\begin{figure}[t!]
	\begin{center}
		\includegraphics[width=0.95\columnwidth]{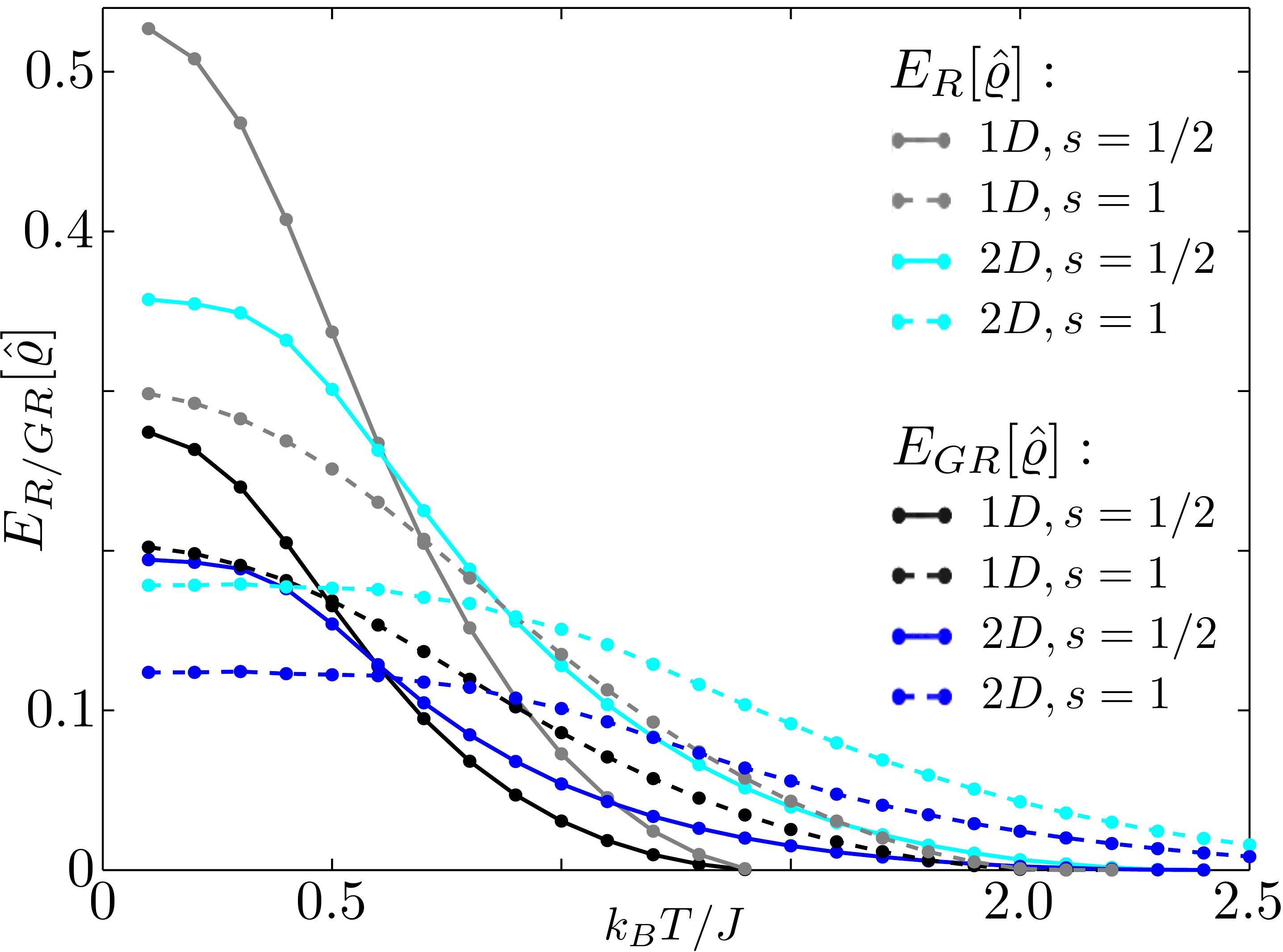}
	\end{center}
	\caption{\label{fig:rob}Lower bounds $E_R[\hat{\varrho}]$ and $E_{GR}[\hat{\varrho}]$ (Eqs.~(\ref{equation:Rbound},\ref{eq:obsRob}), see main text for the choice of the set $Q$) to the robustness measures
$\mathcal{E}_{R}[\hat{\varrho}]$ and $\mathcal{E}_{GR}[\hat{\varrho}]$ as a function of temperature
	for thermal states  of the quasi-one-dimensional (black, see Eqs.~(\ref{hamiltonian:effD},\ref{hamiltonian:heisenberg1D})) and
quasi-two-dimensional (blue, see Eq.~(\ref{hamiltonian:heisenberg1D}))  Heisenberg model with spin number $s=1/2$ (solid) and $s=1$ (dashed). 
Note that for any state $E_R[\hat{\varrho}]\ge E_{GR}[\hat{\varrho}]$.
	 Lines are guides to the eye and $N=900$ spins were simulated \cite{finite_size_footnote}.}
\end{figure}

To present results on the robustness measures $\mathcal{E}_R[\hat{\varrho}]$ and $\mathcal{E}_{GR}[\hat{\varrho}]$, we need to specify
 the set of scattering vectors $Q$ appearing in the lower bounds $E_{R}[\hat{\varrho}]$ and $E_{GR}[\hat{\varrho}]$ in Eqs.~(\ref{equation:Rbound},\ref{eq:obsRob}). 
We use the following choice of scattering vectors:
\begin{equation}
Q(x)=\Bigl\{\vec{q}\in\mathcal{Q}\,\big|\,\sum_{i,j}\me^{\mi \vec{q}(\vec{r}_i-\vec{r}_j)}\sum_\alpha\langle\hat{S}^\alpha_{i}\hat{S}^\alpha_{j}\rangle\ge x\Bigr\}
\end{equation}
and then take $Q$ as the $Q(x)$ that maximizes the lower bound.
In Fig.~\ref{fig:rob}, we present results for all the Heisenberg models that we also considered
for the best separable approximation.

\section{Analytic examples}
\label{section:analytics}
In this section, we discuss resonating valence bond (RVB) states and the Majumdar-Gosh model, for which an exact expression for
the expectation value of $\hat{S}(\vec{q})$ (and hence for our entanglement bounds) may be obtained.

In the context of high temperature superconductors, resonating valence bond (RVB) states were introduced by
Anderson \cite{Anderson1973,Anderson1987}. They are used to describe quantum-spin-liquids, i.e., states without long-range magnetic order
\cite{Balents2010}, and appear as ground states of frustrated antiferromagnets. Such systems and their description by RVB currently receive
increased theoretical as well as experimental attention, see, e.g., Refs.\ \cite{YanHW2011,SchuchPCP-G2012,JiangWB2012,Depenbrock2012} and
Ref.\ \cite{HanHCNR-RBL2012} for a recent neutron scattering investigation of the antiferromagnetic Heisenberg
model on a Kagom\'e lattice.
Besides the characterization of high-$T_c$ superconductors, quantum spin liquids have potential applications for
topological quantum computation \cite{Kitaev2003}. The entanglement properties of RVB states have recently been considered using tools from
quantum information theory \cite{ChandranKSSV2007,PoilblancSP-GC2012}. Consider
a lattice with $N$ (even) sites and a dimer covering $\Delta=\{(i_1,j_1),\dots,(i_{N/2},j_{N/2})\}$, i.e., a collection of pairs of lattice
sites such that each lattice site belongs to exactly one dimer. To any such dimerization, one may associate a valence bond state
$|\psi_{\Delta}\rangle=\otimes_{(i,j)\in\Delta}|\phi_{i,j}\rangle$. Singlet RVB states are superpositions of such states,
$|\psi\rangle=\sum_\Delta c_\Delta|\psi_\Delta\rangle$, where each dimer forms a singlet,
$|\phi_{i,j}\rangle=\frac{1}{
\sqrt{2}}(|\!\!\uparrow\rangle_i|\!\!\downarrow\rangle_j-|\!\!\downarrow\rangle_i|\!\!\uparrow\rangle_j)$. 
The span of all singlet valence bond states is equal to the singlet sector,
i.e., to the spin-zero subspace. 
For these states, in the limit $\vec{q}\rightarrow\vec{0}$, we have
$E_{BSA}[|\psi\rangle\langle\psi|](\vec{q}\rightarrow\vec{0})\ge 1-\frac{1}{NC_s}\sum_\alpha \langle\hat{S}_\alpha^2\rangle=1$, where
$\hat{S}_\alpha=\sum_i\hat{S}_i^\alpha$ the total spin along $\alpha$, i.e.,
\begin{equation}
E_{BSA}[|\psi\rangle\langle\psi|](\vec{q}\rightarrow\vec{0})=1,
\end{equation}
for all $|\psi\rangle=\sum_\Delta c_\Delta|\psi_\Delta\rangle$,
i.e., these states maximally violate the lower bound in Eq.\ \eqref{eq:boundsep} and their entanglement as quantified in terms of the best separable approximation is hence optimally quantified by the neutron scattering observable in Eq.\ \eqref{equation:BSAbound}.

Hamiltonians for which RVB may describe the ground state and explain low-lying excitations include examples with frustration due to
additional next-nearest-neighbor interaction such as the so called Klein Hamiltonian \cite{Klein1982} on
two-dimensional lattices and the Majumdar-Ghosh Hamiltonian in one dimension \cite{MajumdarG1969},
 \begin{equation}
\label{hamiltonian:MG}
\hat H^{MG}_{1D} = 2\sum_{i}  \hat{\vec{S}}_{i} \cdot \hat{\vec{S}}_{i+1} + \sum_{i} \hat{\vec{S}}_{i} \cdot \hat{\vec{S}}_{i+2}.
\end{equation}
In Ref.~\cite{majumdar-ghosh} it was shown that the ratio of nearest-neighbor and
next-nearest neighbor coupling in the quasi-one-dimensional antiferromagnet CuCrO$_4$ is close to $2$, putting this magnet in
the vicinity of the Majumdar-Ghosh point.
Every ground state of $\hat H^{MG}_{1D}$ is a superposition of two two-periodic states given by products of nearest-neighbor
singlets, i.e., a RVB. 
The equal weight ground state may be given explicitly exploiting its description as a matrix product state \cite{mps}. The correlators can
be computed exactly and allow for a particularly concise expression of the structure factor in the thermodynamic limit: The correlators for a single chain of length $L$ are given by
\begin{equation}
\langle \hat{S}^\alpha_{i}\hat{S}^\alpha_{i+r} \rangle = 
\begin{cases}
(-1)^{\frac{L}{2}-1}\frac{(-1)^{r}}{2^{\frac{L}{2}+1}+4(-1)^{\frac{L}{2}}} & \text{ for } r>1, \\
 -\frac{1}{4}\frac{2^{\frac{L}{2}}+4(-1)^{\frac{L}{2}}}{2^{\frac{L}{2}+1}+4(-1)^{\frac{L}{2}}} & \text{ for } r=1.
\end{cases}
\end{equation}
In the thermodynamic limit we find $\langle \hat{S}^\alpha_{i}\hat{S}^\alpha_{i+r} \rangle = -\frac{\delta_{1,r}}{8}$ for $\alpha = x,y,z$ and $r>0$, which
yields
\begin{equation}
E_{BSA}[\hat{\varrho}](\vec{q})=2\cos(q_x)-1
\end{equation}
if every one of the mutually uncoupled chains is in this ground state, see solid line in Fig.~1. With $Q=\{\vec{q}\}$, i.e.,
$|Q|=1$, we find for the robustness bounds in Eq.\ \eqref{equation:Rbound} that, as $N\rightarrow\infty$,
\begin{equation}
\begin{split}
E_R[\hat{\varrho}]&= \frac{1-3\cos(q_x)}{N} -1,\\
E_{GR}[\hat{\varrho}]&= \frac{3(1-\cos(q_x))}{2+N}-1,
\end{split}
\end{equation}
both of which become trivial if $N$ is too large. Just as for the numerical examples, we see that summation over several scattering vectors
is necessary to obtain a non-trivial lower bound: We choose $Q=\{\vec{q}\in\mathcal{Q}\,|\,\frac{2\pi}{L}\frac{L}{2}-\frac{2\pi}{L}cL<
q_x\le \frac{2\pi}{L}\frac{L}{2}+\frac{2\pi}{L}cL \}$, i.e., $|Q|=2cN$. We may then, for large $L$, replace the  summation of the structure factor over different $q_x$ by an integral according to $\lim_{L\rightarrow \infty}\frac{1}{L}\sum_{q=a}^{b} f(\frac{2\pi q}{L}) = \frac{1}{2\pi}\int_{2\pi a/L}^{2\pi b/L} f(q) \mathrm{d} q$. By direct computation of the integral and then maximizing over $0<c<1/2$, we find that
$E_R[\hat{\varrho}]\approx 0.51$ and $E_{GR}[\hat{\varrho}]\approx 0.23$ in the thermodynamic limit.

\section{Summary and outlook}
\label{section:summary}
We showed how entanglement may be quantified relying on observables typically obtained in scattering experiments. In particular, these
observables can be measured via the scattering cross-section in neutron scattering.
We showed how such measurements give lower bounds on the entanglement in the sample, bounding the best separable
approximation, the robustness of entanglement and the generalized robustness of entanglement.
These bounds do neither rely on the knowledge of the systems underlying Hamiltonian nor any other information
about the state of the sample material. The detection can be applied to macroscopic systems,
because the experimental effort does not increase with the system size -- in stark contrast to quantum state tomography.
We showed for several model Hamiltonians such as the Heisenberg, Majumdar-Gosh and XY models (for different spin numbers and different spatial geometries), that our method can
indeed quantify entanglement in large samples at finite temperature. Interestingly, quantum phase transitions and factorization points are detected by our entanglement bounds.
The considered models are well known and applicable to real materials. Therefore our results pave the way for macroscopic entanglement
quantification in experiments. This is very important for future applications, which utilize entanglement, e.g. in quantum information science.
Our method might also be valuable as an alternative way to check the power of a model to describe the sample material, e.g., if a sample is highly entangled,
a classical description certainly fails.

\acknowledgements
We gratefully acknowledge Robert Rosenbach for help with the numerics.
The work at Ulm University has been supported by the EU Integrated Projects Q-ESSENCE and SIQS, the EU STREP EQUAM, the BMBF Verbundprojekt QuOReP, a GIF project, and an Alexander von Humboldt Professorship. 
D.B., M.E. and H.K. acknowledge financial support by the Deutsche Forschungsgemeinschaft (DFG).

\appendix
\begin{widetext}
\section{Bounds for fully separable states}
\label{appendix:bounds}

Let $Q\subset\rr^3$. For $\vec{q}\in Q$, $i=1,\dots,N$, and $\alpha=x,y,z$, let $f_{i,\alpha}(\vec{q})\in\cc$ and $\vec{r}_i\in\rr^3$.
Further, for each $\vec{q}\in Q$ let $M(\vec{q})$ be a $3\times 3$ Hermitian positive semi-definite matrix with
entries $M_{\alpha,\beta}(\vec{q})$. Consider the observable
\begin{equation}
\begin{split}
\hat{S}&=
\sum_{\vec{q}\in Q}\sum_{\alpha,\beta}\sum_{i,j}M_{\alpha,\beta}(\vec{q})f_{i,\alpha}^*(\vec{q})f_{j,\beta}(\vec{q})\me^{\mi\vec{q}(\vec{r}_i-\vec{r}_j)}\hat{S}_i^\alpha\hat{S}_j^\beta,
\end{split}
\end{equation}
which is positive semi-definite: Denoting $\hat{S}_\alpha(\vec{q})=\sum_i
f_{i,\alpha}(\vec{q})\me^{-\mi\vec{q}\vec{r}_i}\hat{S}_i^\alpha$, we have
\begin{equation}
\begin{split}
\hat{S}&=
\sum_{\vec{q}\in Q}\sum_{\alpha,\beta}
M_{\alpha,\beta}(\vec{q})
\hat{S}_\alpha(\vec{q})^\dagger\hat{S}_\beta(\vec{q}),
\end{split}
\end{equation}
which is positive semidefinite as the $M_{\alpha,\beta}(\vec{q})$ are and as for every $\vec{q}$ and every state vector $|\psi\rangle$ the $3\times 3$ matrix with entries $\langle\psi|\hat{S}_\alpha(\vec{q})^\dagger\hat{S}_\beta(\vec{q})|\psi\rangle$ is positive semidefinite.

For each $i=1,\dots,N$ define the $3\times 3$ matrix $M^{(i)}$ with entries
\begin{equation}
M^{(i)}_{\alpha,\beta}=\sum_{\vec{q}\in Q}f_{i,\alpha}^*(\vec{q})f_{i,\beta}(\vec{q})M_{\alpha,\beta}(\vec{q}).
\end{equation}
We assume that these matrices are real symmetric. This is fulfilled, e.g., if $f_{i,\alpha}(\vec{q})=F_i(\vec{q})g_{i,\alpha}$ with $F_i(\vec{q})\in\cc$ and $g_{i,\alpha}\in\rr$. We further note that these $M^{(i)}$ are positive semidefinite as we assumed that the $M_{\alpha,\beta}(\vec{q})$ are positive semidefinite.

We set out to derive upper and lower bounds on the expectation of $\hat{S}$ for product states $|\psi\rangle=\otimes_i|\psi_i\rangle$. The same bounds then also hold for
fully separable states $\hat{\varrho}=\sum_np_n\otimes_i\hat{\varrho}_i^{(n)}$ by convexity.
For product states, we have that for all $i\ne j$ the equality $\langle\hat{S}_i^\alpha\hat{S}_j^\beta\rangle=\langle\hat{S}_i^\alpha\rangle\langle\hat{S}_j^\beta\rangle$ holds. Hence,
\begin{equation}
\label{eq:sep}
\begin{split}
\langle\hat{S}\rangle&=\sum_i\sum_{\alpha,\beta}M^{(i)}_{\alpha,\beta}\langle\hat{S}_i^\alpha\hat{S}_i^\beta\rangle-\sum_i\sum_{\alpha,\beta}M^{(i)}_{\alpha,\beta}\langle\hat{S}_i^\alpha\rangle\langle\hat{S}_i^\beta\rangle\\
&\hspace{2cm}+\sum_{\vec{q}\in Q}\sum_{\alpha,\beta}\sum_{i,j}M_{\alpha,\beta}(\vec{q})f_{i,\alpha}^*(\vec{q})f_{j,\beta}(\vec{q})\me^{\mi\vec{q}(\vec{r}_i-\vec{r}_j)}\langle\hat{S}_i^\alpha\rangle\langle\hat{S}_j^\beta\rangle\\
&=:A-B+C.
\end{split}
\end{equation}

\subsection{Lower bound}
The third term, $C$, in Eq.~(\ref{eq:sep}) is non-negative as for each $\vec{q}$ the matrix $M(\vec{q})$ is positive semidefinite. Hence we have the lower bound
\begin{equation}
\begin{split}
\langle\hat{S}\rangle&\ge A-B=\sum_i\sum_{\alpha,\beta}M^{(i)}_{\alpha,\beta}\left(\langle\hat{S}_i^\alpha\hat{S}_i^\beta\rangle-\langle\hat{S}_i^\alpha\rangle\langle\hat{S}_i^\beta\rangle\right).
\end{split}
\end{equation}
As we assumed that for each $i$ the $M^{(i)}$ are real symmetric, there are mutually orthonormal real eigenvectors $\vec{m}^{(i)}_\gamma$ with corresponding eigenvalues $m_\gamma^{(i)}$ and there is a unitary $\hat{U}_i$ such that $\sum_\alpha [\vec{m}^{(i)}_\gamma]_\alpha\hat{S}_i^\alpha=\hat{U}^\dagger_i\hat{S}_i^\gamma \hat{U}_i$  for all $\gamma$. Thus
\begin{equation}
\begin{split}
\sum_{\alpha,\beta}M^{(i)}_{\alpha,\beta}\left(\langle\psi_i|\hat{S}_i^\alpha\hat{S}_i^\beta |\psi_i\rangle-\langle \psi_i|\hat{S}_i^\alpha |\psi_i\rangle\langle \psi_i|\hat{S}_i^\beta |\psi_i\rangle\right)
&=\sum_{\gamma}m^{(i)}_{\gamma}\left(\langle\psi_i|\hat{U}^\dagger_i(\hat{S}_i^\gamma)^2 \hat{U}_i |\psi_i\rangle-\langle \psi_i|\hat{U}^\dagger_i\hat{S}_i^\gamma \hat{U}_i |\psi_i\rangle^2\right)
\\
&\ge\min_{\substack{|\psi\rangle\in\cc^{2s+1}\\ \langle\psi|\psi\rangle=1}}\sum_{\gamma}m^{(i)}_{\gamma}\left(\langle\psi|(\hat{S}_i^\gamma)^2 |\psi\rangle-\langle \psi|\hat{S}_i^\gamma |\psi\rangle^2\right),
\end{split}
\end{equation}
which is $c_{\min}$ presented in the main text.

\subsection{Upper bound}
We first bound, similar to above,
\begin{equation}
\begin{split}
A&=\sum_i\sum_{\alpha,\beta}M^{(i)}_{\alpha,\beta}\langle\psi_i|\hat{S}_i^\alpha\hat{S}_i^\beta |\psi_i\rangle=
\sum_i\sum_{\gamma}m^{(i)}_{\gamma}\langle\psi_i|\hat{U}^\dagger_i(\hat{S}_i^\gamma)^2 \hat{U}_i |\psi_i\rangle
\le \sum_i\bigl\|\sum_{\gamma}m^{(i)}_{\gamma}(\hat{S}_i^\gamma)^2\bigr\|.
\end{split}
\end{equation}
Now denote by $\mathcal{M}$ the Hermitian $3N\times 3N$ matrix  with entries
\begin{equation}
\begin{split}
\mathcal{M}_{i,\alpha;j,\beta}&=\sum_{\vec{q}\in Q}M_{\alpha,\beta}(\vec{q})f_{i,\alpha}^*(\vec{q})f_{j,\beta}(\vec{q})\left(\me^{\mi\vec{q}(\vec{r}_i-\vec{r}_j)}-\delta_{i,j}\right).
\end{split}
\end{equation}
This matrix has $\text{tr}[\mathcal{M}]=0$, i.e., its largest eigenvalue $\lambda_{\max}$ is non-negative and therefore 
\begin{equation}
C-B=\sum_{\alpha,\beta}\sum_{i,j}\mathcal{M}_{i,\alpha;j,\beta}\langle\hat{S}_i^\alpha\rangle\langle\hat{S}_j^\beta\rangle\le \lambda_{\max}Ns^2.
\end{equation}
Hence, we have the bound
\begin{equation}
A-B+C\le \sum_i\bigl\|\sum_{\gamma}m^{(i)}_{\gamma}(\hat{S}_i^\gamma)^2\bigr\|+ \lambda_{\max}Ns^2.
\end{equation}
This constitutes our general result for $c_{\max}$. To compute it, one needs to find the maximum eigenvalue of the $3N\times 3N$ matrix $\mathcal{M}$
and, for each $i=1,\dots,N$, the eigenvalues of the $3\times 3$ matrix $M^{(i)}$ .
We now discuss a geometry for which this may be made more explicit.

Let the positions of the $i$'th spin be $\vec{r}_i=\vec{r}_{k,l}=\vec{R}_k+\vec{x}_l$, where $k=1,\dots,N_c$ and $l=1,\dots,n$ such that $N=nN_c$. 
Further we let the lattice sites $k=1,\dots,N_c$, with $N_c=N^c_1N^c_2N^c_3$, be the sites of a finite Bravais lattice with primitive vectors
$\vec{a}_d$, $d=1,2,3$, such that
$\vec{R}_k=\sum_{d=1}^3k_d\vec{a}_d$ with $k_d\in\{1,\dots,N^c_d\}$. Note that this is more general than in the main text as we allow for $n$ spins 
in each unit cell. We now assume that $f_{i,\alpha}(\vec{q})=f_{k,l,\alpha}(\vec{q})=f_{l,\alpha}(\vec{q})$, i.e., depends only on $l$.
Further we assume that 
\begin{equation}
Q\subset\Bigl\{\sum_{d=1}^3q_d\vec{b}_d\,\big|\, q_d\in\frac{1}{N^c_d}\{0,\dots,N^c_d-1\}\Bigr\}=:\mathcal{Q},
\end{equation}
where the $\vec{b}_d$ are reciprocal primitive vectors corresponding to the $\vec{a}_d$. We then have $\frac{1}{N_c}\sum_{\vec{p}\in\mathcal{Q}}\me^{\mi\vec{p}(\vec{R}_k-\vec{R}_{k^\prime})}=\delta_{k,k^\prime}$, which yields
\begin{equation}
\begin{split}
\mathcal{M}_{\alpha,k,l;\beta,k^\prime,l^\prime}&=
\sum_{\vec{q}\in Q}M_{\alpha,\beta}(\vec{q})f_{l,\alpha}^*(\vec{q})f_{l^\prime,\beta}(\vec{q})\left(\me^{\mi\vec{q}(\vec{R}_k-\vec{R}_{k^\prime})}\me^{\mi\vec{q}(\vec{x}_l-\vec{x}_{l^\prime})}-\delta_{k,k^\prime}\delta_{l,l^\prime}\right)\\
&=:\sum_{\vec{q}\in Q}M^\prime_{\alpha,l;\beta,l^\prime}(\vec{q})\left(\me^{\mi\vec{q}(\vec{R}_k-\vec{R}_{k^\prime})}-\delta_{k,k^\prime}\delta_{l,l^\prime}\right)\\
&=\sum_{\vec{q}\in Q}M^\prime_{\alpha,l;\beta,l^\prime}(\vec{q})\left(\me^{\mi\vec{q}(\vec{R}_k-\vec{R}_{k^\prime})}-\delta_{l,l^\prime}\frac{1}{N_c}\sum_{\vec{p}\in\mathcal{Q}}\me^{\mi\vec{p}(\vec{R}_k-\vec{R}_{k^\prime})}\right)\\
&=\sum_{\vec{q}\in Q}M^\prime_{\alpha,l;\beta,l^\prime}(\vec{q})\sum_{\vec{p}\in\mathcal{Q}}\left(\delta_{\vec{p},\vec{q}}-\delta_{l,l^\prime}\frac{1}{N_c}\right)\me^{\mi\vec{p}(\vec{R}_k-\vec{R}_{k^\prime})}\\
&=:\sum_{\vec{p}\in\mathcal{Q}}M^{\prime\prime}_{\alpha,l;\beta,l^\prime}(\vec{p})\me^{\mi\vec{p}(\vec{R}_k-\vec{R}_{k^\prime})}\\
&=:\sum_{\vec{p}\in\mathcal{Q}}[M^{\prime\prime}(\vec{p})\otimes\vec{e}_{\vec{p}}\vec{e}_{\vec{p}}^\dagger]_{\alpha,l,k;\beta,l^\prime,k^\prime}
\end{split}
\end{equation}
and thus $\lambda_{\max}=N_c\max_{\vec{p}\in\mathcal{Q}}\lambda_{\max}[M^{\prime\prime}(\vec{p})]=N_c\max_{\vec{p}\in Q}\lambda_{\max}[M^{\prime\prime}(\vec{p})]$, where
\begin{equation}
M^{\prime\prime}_{\alpha,l;\beta,l^\prime}(\vec{p})=\sum_{\vec{q}\in Q}
M_{\alpha,\beta}(\vec{q})f_{l,\alpha}^*(\vec{q})f_{l^\prime,\beta}(\vec{q})\me^{\mi\vec{q}(\vec{x}_l-\vec{x}_{l^\prime})}
\left(\delta_{\vec{p},\vec{q}}-\delta_{l,l^\prime}\frac{1}{N_c}\right).
\end{equation}
Further,
\begin{equation}
M^{(i)}_{\alpha,\beta}=M^{(k,l)}_{\alpha,\beta}=M^{(l)}_{\alpha,\beta}=\sum_{\vec{q}\in Q}f_{l,\alpha}^*(\vec{q})f_{l,\beta}(\vec{q})M_{\alpha,\beta}(\vec{q})
\end{equation}
with eigenvalues $m_\gamma^{(l)}$. We hence have the bound
\begin{equation}
A-B+C\le N_c\sum_{l=1}^n\bigl\|\sum_{\gamma}m^{(l)}_{\gamma}(\hat{S}_i^\gamma)^2\bigr\|+ Ns^2N_c\max_{\vec{p}\in Q}\lambda_{\max}[M^{\prime\prime}(\vec{p})].
\end{equation}
Comparing this $c_{\max}$ to the general bound above, one now has to, for each $\vec{q}\in Q$, find the maximum eigenvalue of a $3n\times 3n$ matrix (where we recall that $n$ is the number of spins in each unit cell) and, for each $l=1,\dots,n$, find the eigenvalues of the $3\times 3$ matrix $M^{(l)}$.

If $f_l(\vec{q})=f(\vec{q})$ and $M_{\alpha,\beta}(\vec{q})=\delta_{\alpha,\beta}/|f(\vec{q})|^2$, we have 
\begin{equation}
\begin{split}
M^{\prime\prime}_{\alpha,l;\beta,l^\prime}(\vec{p})
&=\delta_{\alpha,\beta}\sum_{\vec{q}\in Q}\me^{\mi\vec{q}(\vec{x}_l-\vec{x}_{l^\prime})}\left(\delta_{\vec{p},\vec{q}}-\delta_{l,l^\prime}\frac{1}{N_c}\right)
=\delta_{\alpha,\beta}\me^{\mi\vec{p}(\vec{x}_l-\vec{x}_{l^\prime})}-\delta_{\alpha,\beta}\delta_{l,l^\prime}\frac{|Q|}{N_c}
\end{split}
\end{equation}
and $M^{(l)}_{\alpha,\beta}=|Q|\delta_{\alpha,\beta}$, i.e.,  the bound simplifies to
\begin{equation}
A-B+C\le N|Q|s+ N^2s^2,\label{eq:upperboundsimple}
\end{equation}
which is $c_{\max}$ in the main text.

\section{Simulation details: Effective one- and two-dimensional systems}
\label{appendix:effectiveD}
Consider
\begin{equation}
\label{S_alpha_beta}
\hat{S}_{\alpha,\beta}(\vec{q})=\sum_{\vec{i},\vec{j}}
\me^{\mi\vec{q}(\vec{i}-\vec{j})}
\hat{S}_{\vec{i}}^\alpha\hat{S}_{\vec{j}}^\beta.
\end{equation}
We write $\vec{i}=(i_x\,i_y\,i_z)\in\{1,\dots,L\}^{\times 3}$, $\vec{q}=(q_x\,q_y\,q_z)$, $\tilde{\vec{i}}=(i_x\,i_y)$, $\tilde{\vec{q}}=(q_x\,q_y)$.
If the system consists of mutually uncoupled (in the $z$-direction) two-dimensional systems, we have $\langle
\hat{S}_{\vec{i}}^\alpha\hat{S}_{\vec{j}}^\beta\rangle=\langle\hat{S}_{\vec{i}}^\alpha\rangle\langle\hat{S}_{\vec{j}}^\beta\rangle$ whenever
$i_z\ne j_z$, i.e.,
\begin{equation}
\nonumber
\begin{split}
\langle\hat{S}_{\alpha,\beta}(\vec{q})\rangle&=\sum_{\substack{\vec{i},\vec{j}\\ i_z=j_z}}\me^{\mi\tilde{\vec{q}}\cdot(\tilde{\vec{i}}-\tilde{\vec{j}})}\langle\hat{S}_{\vec{i}}^\alpha\hat{S}_{\vec{j}}^\beta\rangle+\sum_{\substack{\vec{i},\vec{j}\\ i_z\ne j_z}}\me^{\mi\vec{q}\cdot(\vec{i}-\vec{j})}\langle\hat{S}_{\vec{i}}^\alpha\rangle\langle\hat{S}_{\vec{j}}^\beta\rangle\\
&=\sum_{\substack{\vec{i},\vec{j}\\ i_z=j_z}}\me^{\mi\tilde{\vec{q}}\cdot(\tilde{\vec{i}}-\tilde{\vec{j}})}\left(\langle\hat{S}_{\vec{i}}^\alpha\hat{S}_{\vec{j}}^\beta\rangle
-\langle\hat{S}_{\vec{i}}^\alpha\rangle\langle\hat{S}_{\vec{j}}^\beta\rangle\right)
+\Bigl(\sum_{\vec{i}}\me^{\mi\vec{q}\cdot\vec{i}}\langle\hat{S}_{\vec{i}}^\alpha\rangle\Bigr)\Bigl(\sum_{\vec{i}}\me^{\mi\vec{q}\cdot\vec{i}}\langle\hat{S}_{\vec{i}}^\beta\rangle\Bigr)^*\\
&=:S_{\alpha,\beta}(\tilde{\vec{q}})+M_\alpha(\vec{q})M^*_\beta(\vec{q}).
\end{split}
\end{equation}
Now let the two-dimensional sub-systems be equal. Then, any thermal state of the system is of the form $\hat{\varrho}=\otimes_{i_z}\hat{\varrho}_{i_z}$, where the $\hat{\varrho}_{i_z}$ are equal and each describes a two-dimensional layer at $z$ coordinate $i_z$. Hence,
\begin{equation}
\langle\hat{S}_{\vec{i}}^\alpha\rangle=\text{tr}[\hat{S}_{\vec{i}}^\alpha\hat{\varrho}]=\text{tr}[\hat{S}_{\vec{i}}^\alpha\hat{\varrho}_{i_z}]=:\langle \hat{S}_{\tilde{\vec{i}}}^\alpha\rangle_{2D},
\end{equation}
which does not depend on $i_z$. Similarly, for $i_z=j_z$,
\begin{equation}
\langle\hat{S}_{\vec{i}}^\alpha\hat{S}_{\vec{j}}^\beta\rangle=\text{tr}[\hat{S}_{\vec{i}}^\alpha\hat{S}_{\vec{j}}^\beta\hat{\varrho}]=\text{tr}[\hat{S}_{\vec{i}}^\alpha\hat{S}_{\vec{j}}^\beta\hat{\varrho}_{i_z}]=:\langle \hat{S}_{\tilde{\vec{i}}}^\alpha\hat{S}_{\tilde{\vec{j}}}^\beta\rangle_{2D},
\end{equation}
which does not depend on $i_z$. Hence, 
\begin{equation}
\nonumber
\frac{S_{\alpha,\beta}(\tilde{\vec{q}})}{L}=\sum_{\substack{i_x,i_y,\\j_x,j_y}}\me^{\mi\tilde{\vec{q}}\cdot(\tilde{\vec{i}}-\tilde{\vec{j}})}\left(\langle\hat{S}_{\tilde{\vec{i}}}^\alpha\hat{S}_{\tilde{\vec{j}}}^\beta\rangle_{2D}
-\langle\hat{S}_{\tilde{\vec{i}}}^\alpha\rangle_{2D}\langle\hat{S}_{\tilde{\vec{j}}}^\beta\rangle_{2D}\right),
\end{equation}
which does not depend on $q_z$,
and
\begin{equation}
\begin{split}
M_\alpha(\vec{q})&=\sum_{\tilde{\vec{i}}}\me^{\mi\tilde{\vec{q}}\cdot\tilde{\vec{i}}}\langle\hat{S}_{\tilde{\vec{i}}}^\alpha\rangle_{2D}\sum_{i_z}\me^{\mi q_zz_i}
=L\delta_{q_z,0}\sum_{\tilde{\vec{i}}}\me^{\mi\tilde{\vec{q}}\cdot\tilde{\vec{i}}}\langle\hat{S}_{\tilde{\vec{i}}}^\alpha\rangle_{2D},
\end{split}
\end{equation}
where we used that $q_z\in 2\pi\{0,\dots,L\}/L$.
Similarly, if the system is quasi-one-dimensional with $\hat{\varrho}=\otimes_{i_z,i_y}\hat{\varrho}_{i_z,i_y}$ and all the $\hat{\varrho}_{i_z,i_y}$ equal, we have
\begin{equation}
\nonumber
\langle\hat{S}_{\alpha,\beta}(\vec{q})\rangle=S_{\alpha,\beta}(q_x)+M_\alpha(\vec{q})M_\beta(\vec{q})^*,
\end{equation}
where, 
\begin{equation}
\nonumber
\begin{split}
\frac{S_{\alpha,\beta}(q)}{L^2}&=\sum_{i_x,j_x}\me^{\mi q(i_x-j_x)}\left(\langle\hat{S}_{i_x}^\alpha\hat{S}_{j_x}^\beta\rangle_{1D}
-\langle\hat{S}_{i_x}^\alpha\rangle_{1D}\langle\hat{S}_{j_x}^\beta\rangle_{1D}\right),\\
\frac{M_\alpha(\vec{q})}{L^2}&=\delta_{q_z,0}\delta_{q_y,0}\sum_{i_x}\me^{\mi q_xi_x}\langle\hat{S}_{i_x}^\alpha\rangle_{1D}.
\end{split}
\end{equation}

\section{Symmetries}
\label{appendix:symmetries}
\subsection{Heisenberg models}
For all the considered Heisenberg models, we have
$\hat{H}=(\bigotimes_{\vec{i}}\hat{U}_{\vec{i}})\hat{H}(\bigotimes_{\vec{i}}\hat{U}_{\vec{i}})$, where all the $\hat{U}_{\vec{i}}$ implement
the same spin rotation. This implies  $\langle \hat{S}^\alpha_{\vec{i}}\rangle=0$ and $\langle
\hat{S}^\alpha_{\vec{i}}\hat{S}^\beta_{\vec{j}}\rangle=\delta_{\alpha,\beta}\langle \hat{S}^z_{\vec{i}}\hat{S}^z_{\vec{j}}\rangle$. Hence,

\begin{equation}
\begin{split}
E(\vec{q})&=1-\sum_{\alpha,\beta}\tfrac{\delta_{\alpha,\beta}-\bar{q}_\alpha\bar{q}_\beta}{NC_s}\sum_{\vec{i},\vec{j}}
\me^{\mi\vec{q}(\vec{i}-\vec{j})}
\langle\hat{S}_{\vec{i}}^\alpha\hat{S}_{\vec{j}}^\beta\rangle\\
&=1-\tfrac{2}{NC_s}\sum_{\vec{i},\vec{j}}
\me^{\mi\vec{q}(\vec{i}-\vec{j})}
\langle\hat{S}_{\vec{i}}^z\hat{S}_{\vec{j}}^z\rangle.
\end{split}
\end{equation}
For the quasi-one- and two-dimensional system we have
\begin{equation}
E(\vec{q})=1-\tfrac{2}{LC_s}
\sum_{i_x,j_x}\me^{\mi q_x(i_x-j_x)}\langle\hat{S}_{i_x}^z\hat{S}_{j_x}^z\rangle_{1D}
\end{equation}
and
\begin{equation}
E(\vec{q})=1-\tfrac{2}{L^2C_s}
\sum_{\tilde{\vec{i}},\tilde{\vec{j}}}\me^{\mi\tilde{\vec{q}}\cdot(\tilde{\vec{i}}-\tilde{\vec{j}})}\langle\hat{S}_{\tilde{\vec{i}}}^\alpha\hat{S}_{\tilde{\vec{j}}}^\beta\rangle_{2D},
\end{equation}
respectively.

\subsection{XY model}
The Hamiltonian of the quasi-one-dimensional dimensional XY model,
\begin{equation}
\begin{split}
\hat{H} &= \sum_{\langle \vec{i},\vec{j}\rangle}\delta^d_{\vec{i},\vec{j}}
 \bigl[(1+\gamma) \hat S_{\vec{i}}^x \hat S_{\vec{j}}^x + (1-\gamma) \hat S_{\vec{i}}^y \hat S_{\vec{j}}^y \bigr] - h \sum_{\vec{i}} \hat S_{\vec{i}}^z,
\end{split}
\end{equation}
 is invariant under simultaneous rotation of all the spins around their $z$ axis by $\pi$ (which takes $\hat{S}_{\vec{i}}^x$ to $-\hat{S}_{\vec{i}}^x$, $\hat{S}_{\vec{i}}^y$ to $-\hat{S}_{\vec{i}}^y$, and leaves $\hat{S}_{\vec{i}}^z$ invariant), which implies $\langle\hat{S}_{\vec{i}}^x\rangle=\langle\hat{S}_{\vec{i}}^y\rangle=\langle \hat{S}_{\vec{i}}^z\hat{S}_{\vec{j}}^x\rangle=\langle \hat{S}_{\vec{i}}^z\hat{S}_{\vec{j}}^y\rangle=0$. Hence, for $\vec{q}$ such that $\bar{q}_x\bar{q}_y=0$, we find
\begin{equation}
\begin{split}
E(\vec{q})&=1-\sum_{\alpha,\beta}\tfrac{\delta_{\alpha,\beta}-\bar{q}_\alpha\bar{q}_\beta}{NC_s}
\langle\hat{S}_{\alpha,\beta}(\vec{q})\rangle\\
&=1-\tfrac{1}{NC_s}\sum_{\alpha}(1-\bar{q}_\alpha^2)
\langle\hat{S}_{\alpha,\alpha}(\vec{q})\rangle,
\end{split}
\end{equation}
where, assuming $\vec{q}\ne\vec{0}$ and using translational invariance (such that we may write $s_z=\langle\hat{S}_{\vec{i}}^z\rangle$),
\begin{equation}
\begin{split}
\nonumber
\langle\hat{S}_{\alpha,\alpha}(\vec{q})\rangle&=
\sum_{\vec{i},\vec{j}}\me^{\mi\vec{q}\cdot(\vec{i}-\vec{j})}\langle\hat{S}_{\vec{i}}^\alpha\hat{S}_{\vec{j}}^\alpha\rangle
=\sum_{\vec{i},\vec{j}}\me^{\mi\vec{q}\cdot(\vec{i}-\vec{j})}\left(\langle\hat{S}_{\vec{i}}^\alpha\hat{S}_{\vec{j}}^\alpha\rangle-\langle\hat{S}_{\vec{i}}^\alpha\rangle\langle\hat{S}_{\vec{j}}^\alpha\rangle\right)
+\sum_{\vec{i},\vec{j}}\me^{\mi\vec{q}\cdot(\vec{i}-\vec{j})}\langle\hat{S}_{\vec{i}}^\alpha\rangle\langle\hat{S}_{\vec{j}}^\alpha\rangle\\
&=\sum_{\vec{i},\vec{j}}\me^{\mi\vec{q}\cdot(\vec{i}-\vec{j})}\delta_{\vec{i},\vec{j}}^d\left(\langle\hat{S}_{\vec{i}}^\alpha\hat{S}_{\vec{j}}^\alpha\rangle-\langle\hat{S}_{\vec{i}}^\alpha\rangle\langle\hat{S}_{\vec{j}}^\alpha\rangle\right)
+\delta_{\alpha,z}s_z^2N^2\delta_{\vec{q},\vec{0}}\\
&=\sum_{\vec{i},\vec{j}}\me^{\mi q_d(i_d-j_d)}\delta_{\vec{i},\vec{j}}^d\left(\langle\hat{S}_{\vec{i}}^\alpha\hat{S}_{\vec{j}}^\alpha\rangle-\langle\hat{S}_{\vec{i}}^\alpha\rangle\langle\hat{S}_{\vec{j}}^\alpha\rangle\right)\\
&=L^2\sum_{i,j}\me^{\mi q_d(i-j)}\left(\langle\hat{S}_{i}^\alpha\hat{S}_{j}^\alpha\rangle_{1D}-\langle\hat{S}_{i}^\alpha\rangle_{1D}\langle\hat{S}_{j}^\alpha\rangle_{1D}\right)=:L^2\sum_{i,j}c^\alpha_{i-j}(q_d).
\end{split}
\end{equation}
Due to translational invariance, we have $c^\alpha_{l}(q)=c^\alpha_{l+L}(q)=c^\alpha_{l-L}(q)=(c^\alpha_{-l}(q))^*$, and hence for $L$ even,
\begin{equation}
\begin{split}
\sum_{i,j}c^\alpha_{i-j}(q)&=Lc^\alpha_{0}(q)+Lc^\alpha_{L/2}(q)+2L\sum_{l=1}^{L/2-1}\Re[c^\alpha_{l}(q)]
=L\left(\frac{1}{4}-\delta_{\alpha,z}s_z^2\right)+Lc^\alpha_{L/2}(q)+2L\sum_{l=1}^{L/2-1}\Re[c^\alpha_{l}(q)],
\end{split}
\end{equation}
i.e.,
\begin{equation}
\begin{split}
E(\vec{q})&=-1
+4s_z^2(1-\bar{q}_z^2)
-4\sum_{\alpha}(1-\bar{q}_\alpha^2)\left(c^\alpha_{L/2}(q_d)+2\sum_{l=1}^{L/2-1}\Re[c^\alpha_{l}(q_d)]
\right).
\end{split}
\end{equation}
For the correlations functions, we use the results of \cite{BarouchMcC1971} ($1\le l\le L/2$), 
\begin{equation}
\me^{-\mi ql}c^x_l(q)=\frac{1}{4}\left|
\begin{array}{cccc}
G_{-1}&G_{-2}&\cdots&G_{-l}\\
G_0&G_{-1}&\cdots &G_{-l+1}\\
\vdots&\vdots &\ddots &\vdots\\
G_{l-2}&G_{l-3}&\cdots &G_{-1}
\end{array}
\right|,\;\;\;
\me^{-\mi ql}c^y_l(q)=\frac{1}{4}\left|
\begin{array}{cccc}
G_{1}&G_{0}&\cdots&G_{-l+2}\\
G_2&G_{1}&\cdots &G_{-l+3}\\
\vdots&\vdots &\ddots &\vdots\\
G_{l}&G_{l-1}&\cdots &G_{1}
\end{array}
\right|,\;\;\;\me^{-\mi ql}c^z_l(q)=-\frac{1}{4}G_lG_{-l},
\end{equation}
where, for $L\rightarrow\infty$,
\begin{equation}
\begin{split}
s_z&=\frac{1}{2\pi}\int_0^\pi\md\phi\, \frac{\text{tanh}[\beta\Lambda(\phi)/2]}{\Lambda(\phi)}[h-\cos(\phi)],\\
G_l&=\frac{1}{\pi}\int_0^\pi\md\phi\,\frac{\text{tanh}[\beta\Lambda(\phi)/2]}{\Lambda(\phi)}\cos(\phi l)[h-\cos(\phi)]
+\frac{\gamma}{\pi}\int_0^\pi\md\phi\,\frac{\text{tanh}[\beta\Lambda(\phi)/2]}{\Lambda(\phi)}\sin(\phi l)\sin(\phi),\\
\Lambda(\phi)&=\sqrt{\gamma^2\sin^2(\phi)+(h-\cos(\phi))^2}.
\end{split}
\end{equation}

\end{widetext}

\begin{thebibliography}{99}
\bibitem{finite_size_footnote} We have studied systems with total number of spins varying between $100$ and $2500$ and found no noticeable finite-size effects (differences between the obtained bounds were smaller than $0.01$).
\end{thebibliography}
\end{document}